\documentclass[3p,times,procedia]{elsarticle}

\usepackage{ecrc}

\volume{00}

\firstpage{1}

\journalname{Procedia Computer Science}

\runauth{}

\jid{procs}

\jnltitlelogo{Procedia Computer Science}

\CopyrightLine{2011}{Published by Elsevier Ltd.}

\usepackage{amssymb}

\usepackage[figuresright]{rotating}
\usepackage{amssymb}
\usepackage{graphicx}
\usepackage{float}
\usepackage{amsmath}
\usepackage{subfigure}
\usepackage{rotating}
\usepackage{multirow}
\usepackage{array}
\usepackage{algorithm}
\usepackage{algorithmic}
\usepackage{url}
\urldef{\mailsa}\path|{nadeemjvaid@yahoo.com, klatif007@yahoo.com, ashfaqahmad@yahoo.com},|

\begin{document}

\begin{frontmatter}
\dochead{4th International Conference on Ambient Systems,\\ Networks and Technologies (ANT), 2013}

\title{Divide-and-Rule Scheme for Energy Efficient\\ Routing in Wireless Sensor Networks}

\author{K. Latif$^{\ddag}$, A. Ahmad$^{\ddag}$, N. Javaid$^{\ddag}$, Z. A. Khan$^{\$}$, N. Alrajeh$^{\sharp}$}

\address{$^{\ddag}$COMSATS Institute of Information Technology, Islamabad, Pakistan. \\
        $^{\$}$Faculty of Engineering, Dalhousie University, Halifax, Canada.\\
        $^{\sharp}$B.M.T., C.A.M.S, King Saud University, Riyadh, Saudi Arabia.\\
}

\begin{abstract}
From energy conservation perspective in Wireless Sensor Networks (WSNs), clustering of sensor nodes is a challenging task. Clustering technique in routing protocols play a key role to prolong the stability period and lifetime of the network. In this paper, we propose and evaluate a new routing protocol for WSNs. Our protocol; Divide-and-Rule (DR) is based upon static clustering and dynamic Cluster Head (CH) selection technique. This technique selects fixed number of CHs in each round instead of probabilistic selection of CH. Simulation results show that DR protocol outperform its counterpart routing protocols.
\end{abstract}


\end{frontmatter}

\section{Background}
Spatially dispersed wireless sensor nodes and one or more Base Stations (BSs) are embodied to form WSN. Sensor nodes keep an eye on the physical or environmental conditions at different locations, and communicate efficiently with BS. Generally BS is power rich and nodes are equipped with low power. Applications of WSNs are in security, traffic management, environment monitoring, medical applications, surveillance, etc.

Today's research challenge in WSNs is coping with low power communication. Routing protocols in this regard plays a key role in efficient energy utilization. In sending data from node to BS, selection of a specific route, which tend to minimize the energy consumption is necessary. Old fashioned routing techniques are not as energy efficient as present day clustering techniques. LEACH [1], LEACH-Centralized [2] and Multihop-LEACH [3] are few of the earlier techniques of cluster based routing protocols for WSNs. Basically two types of clustering techniques exist; static clustering and dynamic clustering. Clusters once established and never be changed throughout network operation are known as static clusters, while clusters based on some sort of network characteristics and are changing during network operation are known as dynamic clusters.

Proposed DR scheme is based on static clustering and minimum distance distance based CH selection. Network area is logically divided into small regions (clusters). These regions are abbreviated as NCR1, NCR2, NCR3, etc, as shown in figure 1. Nodes in each region select a CH except the region closest to the BS, that is, region; R1. Nodes whose coordinates lie within the perimeter of R1, communicates directly with BS. Selection of CHs in rest of the regions are based on reference point in each region; reference point is the mid point of each region. Node closest to reference point is selected as CH first, then next closest node and so on till least closest node. In each round only one CH is selected in each region furthermore, we uses multi-hop technique for inter region communication to reduce communication distance. DR scheme has the ability to select CH independent of random number and minimize communication distance to almost less than or equal to reference distance. DR scheme uses hybrid theme of static clustering and dynamic CH selection. This technique divides whole network area into $4(n-1)$ Corner Regions (CRs) and $4(n-1) + 1$ Non Corner Regions (NCRs). CHs are selected from NCRs only. Nodes of central region (NCR1) communicates directly with BS while, nodes of CRs associate with adjacent side neighbour CH. DR scheme minimizes communication distance, prolong stability period, enhances network lifetime, and increases throughput.

Now a day in many application, it is needed that sensor nodes are location aware. The Localization problem is discussed in [4] and [5]. Location awareness of sensor nodes help in removing coverage holes and movement of new sensor nodes in place of dead nodes. Energy holes in sensor networks also causes depletion of network energy quickly. The analysis and modeling of energy hole of different routing protocols are discussed in [6] and [7].

\section {Proposed Scheme}
Localization problem is commonly addressed by many researchers. In localization, network field area is logically divided into sub areas [4], [5]. This may helps in controlling the coverage hole. In our technique we divide the network field into sub regions. The complete formation of region's and detailed operation of our scheme is discussed in this section.

\subsection{Formation of Regions }
In traditional cluster formation technique, CHs are elected on probabilistic bases and threshold calculated for each CH. Nodes then associate with each CH based upon received signal strength thus, forming a cluster. In our protocol we divide entire network area into small logical regions. The division of the regions is such that it reduces the communication distance between node to CH and CH to BS. Following two steps describes formation of regions in detail.

In first step network is divided into $n$ equal distant concentric squares. For simplicity, we take $n=3$ here therefore, network is divided into three equal distance concentric squares: Internal square($I_s$), Middle square($M_s$) and Outer square($O_s$). BS is located in the centre of network field therefore, its coordinates are taken as reference point for formation of concentric squares. Division of network field into concentric squares can be obtained from following equations.

Coordinates of top right corner of $I_s$, $T_r(I_s)$.
\begin{equation}
T_r(I_s) = (C_p (x) + d, C_p(y) + d)    
\end{equation}

Coordinates of bottom right corner of $I_s$, $B_r(I_s)$
\begin{equation}
B_r(I_s) = (C_p (x) + d, C_p(y) - d)  
\end{equation}

Coordinates of top left corner of $I_s$, $T_l(I_s)$
\begin{equation}
T_l(I_s) = (C_p (x) - d, C_p(y) + d)    
\end{equation}

Coordinates of bottom left corner of $I_s$, $B_l(I_s)$
\begin{equation}
B_l(I_s) = (C_p (x) - d, C_p(y) - d)    
\end{equation}

Where, $d$ is the factor of distance from center of network to boundary of $I_s$. value of $d$ for $M_s$ and $O_s$ increases with a multiple of 2 and 3 respectively. If we have $n$ number of concentric squares then we can find the coordinates of $n^{th}$ square, $S_n$ from the following equations. 
%
%

\begin{equation}
T_r(S_n) = (C_p(x) + d_n , C_p(y) + d_n),  
\end{equation}

\begin{equation}
B_r(S_n) = (C_p(x) + d_n , C_p(y) - d_n),  
\end{equation}

\begin{equation}
T_l(S_n) = (C_p(x) - d_n , C_p(y) + d_n), and  
\end{equation}

\begin{equation}
B_l(S_n) = (C_p(x) - d_n , C_p(y) - d_n).  
\end{equation}

In second step we divide the area between two concentric squares into equal area quadrilaterals; latter we name them as Corner  Regions (CR) and Non Corner  Regions (NCR). To divide area between $I_s$ and $M_s$ into four equal area quadrilaterals, we take the top right and bottom right corners of $I_s$ as the reference points. Adding factor $d$ in the x-coordinate of top right and bottom right corner of $I_s$, i.e., $T_r(I_s(x+d, y))$ and $B_r(I_s(x+d, y))$, we get the co-ordinates of NCR2. Adding factor $d$ in the y-coordinate of top right and top left corner of $I_s$, i.e., $T_r(I_s(x, y+d))$ and $T_l(I_s(x, y+d))$, we get the co-ordinates of region NCR3. Subtracting factor $d$, in the x-coordinate of top left and bottom left corner of $I_s$, i.e., $T_l(I_s(x-d, y))$ and $B_l(I_s(x-d, y))$, we get the co-ordinates of region NCR4. Subtracting factor d, in the y-coordinate of bottom right and bottom left corner of $I_s$, i.e., $B_r(I_s(x, y-d))$ and $B_l(I_s(x, y-d))$, we get the co-ordinates of region NCR5. Remaining areas left are the four CR, that is, CR2, CR3, CR4, CR5.
Following the same sequence,  we can divide the area between $M_s$ and $O_s$ into four equal area quadrilateral regions (NCR6, NCR7, NCR8, NCR9)  and corner regions (CR6, CR7, CR8, CR9), as shown in figure 1.

\begin{figure*}[ht!]       
\centering
\includegraphics[height=6cm,width=6cm]{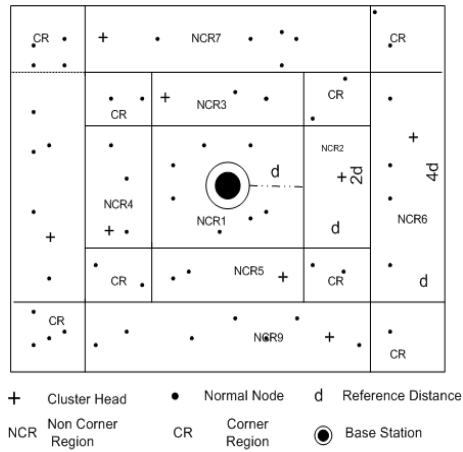}
\caption{Region's formation}
\label{DRlayout4}
\end{figure*}

\subsection{CH Selection}
DR protocol considers multi-hop communication for inter-cluster communication. As we assume n=3 therefore, inter-cluster communication is performed at two levels, that is, at primary level and at secondary level. Our CH selection follows, following approach.

\subsubsection{Primary Level CH}
Primary level CH selection follows the sequence; (i) nodes whose co-ordinates lie in ($I_s$) are nearer to BS therefore, they send data directly to BS, (ii) as clusters are static, therefore one CH is selected in each NCR, (iii) mid point of each NCR is considered as reference point for selection of CH in that region, (iii) nearest node from central reference point is selected as CH and, (iv) next nearest node from the reference point is selected as CH for next round and so on.

\subsubsection{Secondary Level CH.}
Steps followed in selection of secondary level CHs are; (i) CHs in $O_S$ regions, send data to CHs of exactly one level above adjacent region's CH. These CHs are also known as secondary level CHs, (ii) secondary level CHs aggregate their own cluster nodes data and, data of the primary level CH then, transmit data to BS.
\subsection{Protocol Operation}

In setup phase BS divides the network field into small regions, on the bases of their co-ordinates. $I_s$ nodes send data directly to BS. In each region one CH is selected per round. CHs of $O_s$ regions, select front neighboring CHs of $M_S$ regions as their next hop CH. Nodes of CR selects, BS or neighbouring CHs as their CH, based on minimum distance. If a tie occurs, for a node of CR, in selection of CH from its neighbouring regions than, it is resolved by selecting the CH with greater residual energy.

In steady state phase each node send its data to CH in its allocated time slot. Primary level CHs send aggregated data to their respective secondary level CHs. Secondary level CHs then, aggregate all collected data and forward it to BS.
%
\section{Energy Consumption Model}
In this section, we develop a mathematical model, which describes how energy is consumed in different regions of the network field. Basic energy consumption model is adopted from [8]. Equation \ref{radio1} \& \ref{radio2} adopted from [8] shows energy cost of transmission, $T_{Eenergy}$ and reception, $R_{Energy}$ respectively for 1-bit of data over distance $D$ meters.
\begin{equation}\label{radio1}
  T_{Eenergy} = \begin{cases} E_{elec} + \varepsilon _{f_{s}} D^{2} & \mbox{if } D < D_{0} \\ E_{elec} + \varepsilon _{amp} D^{4}, & \mbox{if } D \geq D_{0} \end{cases}
\end{equation}
\begin{equation}\label{radio2}
  R_{Energy} = (E_{elec} )
\end{equation}
\subsection{Energy Consumption in $I_s$}
Following equation calculates the area and energy consumption of $I_{s}$.
\\From figure \ref{DRlayout4}, each side of $I_{s}$ is $2d$ in length and width therefore, Area of $I_{s}$, $A(I_{s})$ :

\begin{equation}
 A(I_{s}) = 4d^{2}
\end{equation}
 Number of nodes in $I_{s}$, $N(I_{s})$ :
\begin{equation}
  N(I_{s}) = 4\rho d^{2}
\end{equation}
where $\rho$ is the node density per unit area.
  Nodes of $I_{s}$ transmit data directly to BS therefore, their energy consumption, $E^{T_{x}}_{I_{s}}$ is given by following equation.
\begin{equation}
  E^{T_{x}}_{I_{s}} = 4\rho d^{2} T_{Energy}
\end{equation}

\subsection{Energy Consumption in CRs}
From figure \ref{DRlayout4}, $d$ is the reference distance for the formation of NCR. Therefore Area of CR, $A(CR)$ is given by:
\begin{equation}
 A(CR) = d^{2}
\end{equation}
 Number of nodes, $N(CR)$ in CR :
\begin{equation}
  N(CR) = \rho d^{2}
\end{equation}
 Nodes of CR may transmit data to BS or to neighbouring NCR's CH, depending on the minimum distance therefore, their energy consumption, $E^{T_{x}}_{CR}$ for sending data to BS is given by equation:
\begin{equation}
  E^{T_{x}}_{CR} = (1-\emph{P})\rho d^{2} T_{Energy}
\end{equation}
\\ where \emph{P} is the probability of sending data to CH, and $(1-\emph{P})$ is the probability of sending data to BS.
\subsection{Energy Consumption in $M_{s}$}
First we calculate energy consumption of normal nodes. Area of each NCR in $M_{s}$ is $2d{^2}$. There are four NCRs and four CRs. Each CR node may associate with one of the NCR's CH or send data directly to BS. Energy consumption of normal nodes in $M_{s}$, per NCR, $E^{T_{x}}_{M_{s}/NCR}$ is given by following equation.
\begin{equation}
  E^{T_{x}}_{M_{s}/NCR} = (2\rho  d^{2}-1)T_{Energy}
\end{equation}
Now we calculate energy consumption of CHs. There are total four CHs in four regions. Each CH consumes energy in transmit ($E^{T_{x}}_{M_{s}\_CH}$), aggregation ($\phi$) and receive ($E^{Rx}_{M_{s}\_CH}$) process therefore, their energy consumption is calculated individually.
\begin{description}
  \item[Transmit Energy]
    \begin{equation}\label{Transmit Energy}
     E^{T_{x}}_{M_{s}\_CH} = (2\rho  d^{2} + P\rho d^{2})T_{Energy} + \phi
    \end{equation}
  Transmit energy of all CHs, $E^{T_{x}}_{M_{s}\_all\_CH}$ in region's of $M_s$ is given by following equation :
    \begin{equation}
     E^{T_{x}}_{M_{s}\_all\_CH} = (8\rho  d^{2} + 4P\rho d^{2})T_{Energy} + 4\phi
    \end{equation}

  \item[Receive Energy]
   \begin{equation}
     E^{Rx}_{M_{s}\_CH} = ((2\rho  d^{2} -1) + P\rho d^{2}) R_{Energy}
   \end{equation}
   Receive energy of all CHs, $E^{Rx}_{M_{s}\_all\_CH}$ in regions's of $M_s$ is given by following equation :
    \begin{equation}
         E^{Rx}_{M_{s}\_all\_CH} = ((8\rho  d^{2} -4) + 4P\rho d^{2})R_{Energy}
    \end{equation}

\end{description}
Total energy consumed in region's of $M_s$, $E^{Tot}_{M_{s}}$ is given by following equation :
\begin{equation}
  E^{Tot}_{M_{s}} = E^{T_{x}}_{M_{s}\_node} + E^{T_{x}}_{M_{s}\_all\_CH} + E^{Rx}_{M_{s}\_all\_CH}
\end{equation}
\subsection{Energy Consumption in $O_{s}$}
In DR protocol, the area of each NCR increases from inner to outer square. The area of NCR of outer square region increases in the following fashion.

Length of one side of $M_s$'s, NCR = $2d$. Length of one side of $O_s$'s, NCR = $4d$ and so on. Width of all the regions remains same that is $d$. Considering this value of length and width into account, area of each NCR of the $O_s$ can be calculated as $4d^{2}$. Taking the area into account we can calculate the total energy consumption, $E^{Tot}_{O_{s}}$ of the $O_s$ from the following equation.

\begin{equation}
  E^{Tot}_{O_{s}} = E^{T_{x}}_{O_{s}\_node} + E^{T_{x}}_{O_{s}\_all\_CH} + E^{Rx}_{O_{s}\_all\_CH}
\end{equation}

\section{Performance Evaluation}        
\subsection{Network Model}
We evaluate our proposed DR scheme by comparing it with LEACH and LEACH-C. We assume a network model of $100$ nodes, with homogeneous initial energy of nodes, which are randomly deployed in each region of  $100 \times 100 m^2$ network area. BS is assumed at the centre of network area. Interference effects in wireless channels are ignored. For simulation purpose, we used MATLAB simulator and first order radio model parameter's are assumed, as shown in table 2.
\begin{table}[h]
\centering
\caption{Radio parameters}
\begin{tabular}{|l|l|}
\hline
Operation & Energy Dissipated  \\
\hline
Transmitter / Receiver Electronics  & Eelec=Etx=Erx=50nJ/bit \\
\hline
Data aggregation energy & EDA=5nJ/bit/signal \\
\hline
Transmit amplifier (if d to BS$<$do) & Efs=10pJ/bit/$4m^{2}$ \\
\hline
Transmit amplifier (if d to BS$>$do) & Emp=0.0013pJ/bit/$m^{4}$\\
\hline
\end{tabular}
\end{table}

\subsection{Results}        
In this section, we evaluate our proposed protocol in terms of stability period, network life time and throughput. Average results obtained after 50 times execution of our protocol.

\begin{figure}[t]       
\centering
\includegraphics[height=7cm,width=10cm]{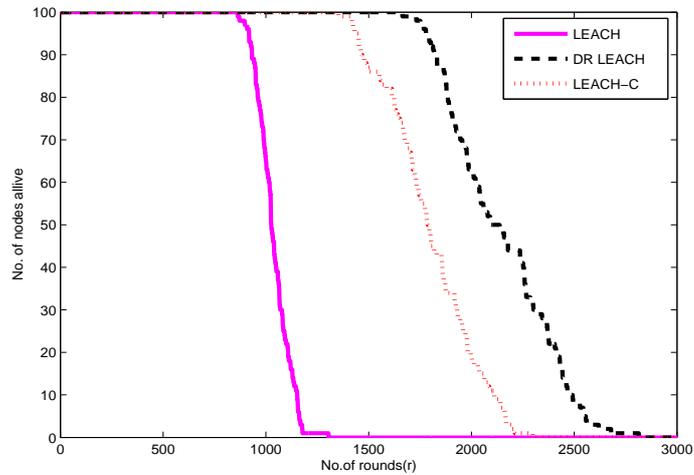}
\caption{Comparison: Rate of Alive Nodes}
\label{drAllive}
\end{figure}


\begin{figure}[H]       
\centering
\includegraphics[height=7cm,width=10cm]{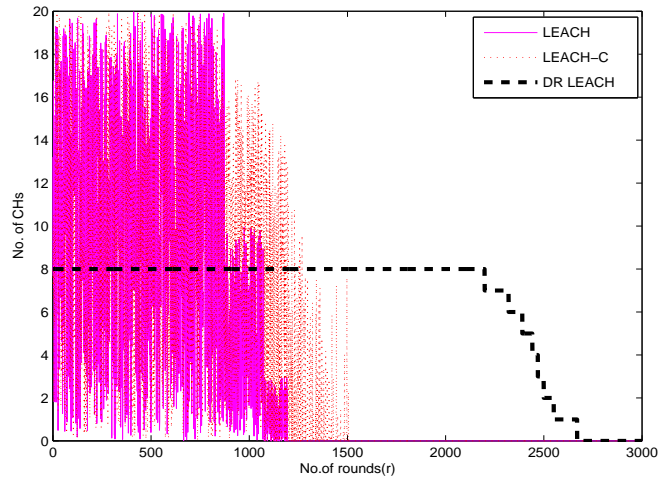}
\caption{Comparison: Rate of CHs per round}
\label{DrChs}
\end{figure}

\subsection{Stability Period}    
Here we evaluate the performance of DR in terms of stability period by comparing it with LEACH and LEACH-C. Figure \ref{drAllive} shows that, our proposed protocol carry out maximum rounds till the death of first node. DR perform 28.63$\%$ better than LEACH and 12.31$\%$ better than LEACH-C. The reason is straight forward; distant nodes of CRs are not enforced to associate with CH. Nodes of CRs may associate either with minimum distant CH or minimum distant BS. Thus, DR minimize communication distance. DR selects optimal number of CHs.

\subsection{Number of CHs per Round}
Figure \ref{DrChs} shows number of CHs formed in each round are fixed. Which shows that near to optimum number is achieved and load is balanced throughout the network operation time, a step towards efficient energy utilization.

\section{Conclusion and Future Work}        
In this paper we have proposed a new clustering techniques for ad-hoc WSNs. DR Scheme uses static clustering and minimum distance based CH selection. We have used a two level hierarchy for inter cluster communication. The beauty of our technique is the formation of square and rectangular regions, which divides the network field into small regions, as a result the communication distance for intra cluster and inter cluster reduces. However CR nodes associate with CH or BS depending on the minimum distance. Our proposed DR scheme uses a hybrid approach of static clustering and dynamic CH selection.
In MATLAB simulation we compared our results with LEACH and LEACH-C. Characteristics of achieving optimum number of CHs in each round and hierarchical inter CHs communication of our technique provided better results than its counterparts, in terms of  stability period, network life time, area coverage and throughput. However, Large network area and greater number of nodes decrease DR efficiency in terms of energy consumption. Another drawback arises when Cluster members associate with CH of its own region even if CH of other region is at a shorter distance. In future we would like to compensate deficiencies explained in this section and implementation of DR in clustering protocols like Threshold sensitive energy efficient sensor network protocol [9], stable election protocol [10], distributed energy efficient clustering [11], etc.


\end{document}